\documentclass{ws-ijmpd}
\usepackage{graphicx}

\def\ut#1{\mathop{\vtop{\ialign{##\crcr
     $\hfil\displaystyle{#1}\hfil$\crcr\noalign
     {\kern1pt\nointerlineskip}\hbox{$\hfil\sim\hfil$}\crcr
     \noalign{\kern1pt}}}}}

\def\undersymbol#1#2{\mathop{\vtop{\ialign{##\crcr
     $\hfil\displaystyle{#2}\hfil$\crcr\noalign
     {\kern1pt\nointerlineskip}\hbox{$\hfil#1\hfil$}\crcr
     \noalign{\kern1pt}}}}}
\def\arcsec{^{\prime\prime}}

\begin{document}

%

\def\nocropmarks{\vskip5pt\phantom{cropmarks}}


%


%
\catchline{}{}{}
%

\title{GRAVITATIONAL WAVE SCINTILLATION BY A STELLAR CLUSTER}


\author{G. Congedo, F. De Paolis, P. Longo, A. A. Nucita, D. Vetrugno}
\address{Department of Physics and INFN, University of Lecce, CP 193,
I-73100 Lecce, Italy}
\author{Asghar Qadir}
\address{Centre for Advanced Mathematics and Physics at the National
University of Science and Technology, Rawalpindi, Pakistan\\
Department of Mathematical Sciences, King Fahd University of
Petroleum and Minerals, Dhahran 31261, Saudi Arabia}


\maketitle

\pub{Received (received date)}{Revised (revised date)}

\begin{abstract}
The diffraction effects on gravitational waves propagating through a
stellar cluster are analyzed in the relevant approximation of
Fresnel diffraction limit. We find that a gravitational wave
scintillation effect - similar to the radio source scintillation
effect - comes out naturally, implying that the gravitational wave
intensity changes in a characteristic way as the observer moves.
\end{abstract}





\section{Introduction}

General Relativity predicts the existence of gravitational waves
that propagate in the vacuum with the speed of light. Analogously to
electromagnetism, it is expected that gravitational waves might also
be scattered by intervening particles \cite{stm,tsm,macquart} and
micro-lensed by compact objects along the line of sight
\cite{ruffa,dinq,din}. Moreover, gravitational waves will  be
diffracted as they pass through a distribution of objects, each of
which acts as an obstacle (or slit). Of course, these effects are
expected to become apparent only if the linear sizes of the
intervening particles are comparable to the gravitational radiation
wavelength in the range $10^{7}-10^{15}$ cm. We consider a stellar
cluster as the diffracting object and concentrate in particular on
that possibly hosted in the center of the Milky Way and surrounding
the central black hole Sgr A$^*$.

There are two types of diffraction phenomena discussed: the
Fraunhoffer \cite{bre} and Fresnel approximation. In our case only
the second is relevant since the first approximation involves
infinite source-cluster and cluster-observer distances. We determine
the conditions under which the diffraction effects may be
measurable.

It turns out that a gravitational wave scintillation effect,
analogous to the well known scintillation of radio sources, comes
out naturally implying that the gravitational wave intensity
changes in a characteristic way with the motion of the observer.

The paper is structured as follows: in Section 2 we describe the
parameters of the cluster mass density profile used to study the
Fresnel diffraction. In Section 3 is given the mathematical tools
of the Fresnel theory, in the particular geometry described there.
In Section 4 we present the main result of this paper: namely, the
scintillation pattern of the gravitational waves as they pass
through the stellar cluster at the galactic center. Finally, in
Section 5 we present some conclusions.

\section{The Mass Distribution Model}

The ESO and Keck teams \cite{Ghez04,Ghez05} have continuously
monitored the region of the galactic center for several years so
that a cluster of stars (at a distance $< 1\arcsec $) has been
identified. In particular, Ghez et al. \cite{Ghez03} have reported
on the observations of the main sequence S2 star (with mass $M_{\rm
S2}\simeq 15$ M$_{\odot}$) orbiting the central black hole with a
Keplerian period of $\simeq 15$ yrs. This has allowed \cite{Ghez05}
the mass contained within $R\simeq4.87\times 10^{-3}$ pc to be
constrained to $M(R)\simeq 3.67\times 10^6 ~M_{\odot}$. The most
plausible model is that this mass is in the form of a super-massive
black hole.

However, there is a possibility that a small fraction of the
inferred central mass is in the form of a diffuse stellar cluster
which surrounds the central black hole. In this case, according to
the dynamical observations towards the galactic center, we require
that the total mass satisfies the condition
$M(R)=M_{BH}+M_{CL}(R)$, $M_{BH}$ and $M_{CL}(R)$ being the black
hole and the cluster mass within $R$, respectively.

Of course, the cluster mass and density distribution are still
unknown so that, as a toy model, we can assume that the stellar
component follows a Plummer density profile given by
\begin{equation}
\rho_{CL}(r) = \rho_0 f(r)~,~~~~~~~~{\rm
with}~~~~~~~~~~f(r)={\left[1+\left(\frac{r}{r_c}\right)^2\right]^{-5/2}}~,\label{plummer1}
\end{equation}
where the cluster central density $\rho _0$ is given by
\begin{equation}
\rho _0 = \frac{M_{CL}}{\int _0^{R_{CL}} 4\pi r^2 f(r)~dr}~,
 \label{plummer2}
\end{equation}
$R_{CL}$ being the cluster radius. Useful information is provided by
the black hole mass fraction, $\lambda_{BH}=M_{BH}/M$ and its
complement, $\lambda_{CL}=1-\lambda_{BH}$. As one can see, the
requirement given in eq. (\ref{plummer2}) implies that
$M(r)\rightarrow M_{BH}$ for $r\rightarrow 0$. Hence, the total mass
density profile $\rho(r)$ is given by
\begin{equation}
\rho(r) = \lambda_{BH} M \delta ^{(3)}(\overrightarrow{r}) +\rho_0
f(r)~ \label{totaldensity}
\end{equation}
and the mass contained within $r$ is
\begin{equation}
M(r) = \lambda_{BH}M + \int_0^r 4\pi r'^2\rho_0 f(r')~dr'~.
\end{equation}
Considerations on both the dynamics and evolution of the stellar
cluster surrounding the central black hole allow us to constrain
the possible range of values \cite{s2} for $\lambda_{BH}$ to be
above about 0.9. In the following for definiteness we will assume
that $\lambda_{BH}\simeq 0.99$.

Assuming that all the cluster components have the same mass
($\simeq 1$ M$_{\odot}$), we can estimate the average distance
between two neighboring stars as $\langle d (r) \rangle \simeq
[{\rm M}_{\odot}/\rho(r)]^{1/3}$.

\section{The Fresnel Diffraction Approximation}

Consider a distant source emitting gravitational radiation. Assume
that there is a distribution of objects between the source and the
observer, each of which acts as an obstacle (or slit). Let
$(X_0,Y_0)$, $(X_1,Y_1)$ and $(X_2,Y_2)$ be three reference frames
(with mutually parallel axes) in the planes of the observer,
diffraction and source, respectively chosen such that the origins
are collinear.

Using the Fresnel approximation for interaction of electromagnetic
waves with dust particles distributed along the line of sight
gives \cite{born,moniez} the diffraction amplitude $A_0$ on the
plane of the observer (at distance $z_0$ from the diffraction
plane) is
\begin{equation}
\label{}
A_0(x_0,y_0) = \frac{e^{i\kappa z_0}}{i\lambda z_0} \int\int^{\infty}_{-\infty}
A'_1(x_1,y_1)e^{\frac{i\kappa}{2z_0}[(x_0 - x_1)^2 + (y_0 - y_1)^2]}dx_1dy_1,\\
\end{equation}
where
\begin{equation}
\label{a1} A'_1(x_1,y_1) = \frac{A_2e^{i\kappa
r_{12}}}{z_1}e^{i\kappa \delta(x_1,y_1)} \label{A}
\end{equation}
is the amplitude just past the diffraction plane. In eq. (\ref{A})
$A_2$ is the source amplitude, $r_{12}\simeq z_1$ is the distance
of the source from a generic point on the diffraction plane, and
$z_1$ is the distance of the source from the center of the
diffraction plane. It is worth noting that when the wave moves
through the diffraction plane, a change of phase does occur due to
the optical path difference $\delta(x_1,y_1)$ which, in turn,
depends on the coordinates of each obstacle. For incoming plane
waves generated by a point-like source, the factor
$A=A_2e^{i\kappa r_{12}}/z_1$ is a real constant so that the
previous integral becomes
\begin{equation}\label{ampiezza}
A_0(x_0,y_0) = \frac{Ae^{i\kappa z_0}}{2i\pi R^2_F}
\int\int^{\infty}_{-\infty} e^{i\kappa\delta(x_1,y_1)}
e^{i\frac{(x_0 - x_1)^2 + (y_0 - y_1)^2}{2R^2_F}}dx_1dy_1,\\
\end{equation}
where the Fresnel radius is defined as
\begin{equation}
R_F=\sqrt{\frac{\lambda z_0}{2\pi}}.
\end{equation}

It is assumed \cite{moniez,vetrugno} that the screen is a step of
optical path $\delta$ parallel to the $Y_1$ axis, described by a
Heaviside distribution in the $(X_1,Y_1)$ plane
$\delta(x_1,y_1)=\delta \times H(x_1)$. This case is realistic
since, at the Fresnel scale, the edge of a particle distribution
can be considered as a straight line that divides the plane into
two regions. Under these assumption, the integral in eq.
(\ref{ampiezza}) turns out to be
\begin{equation}
\label{} A_0(x_0,y_0) = Ae^{i\kappa z_0}\bigg[1 +
\frac{e^{i\kappa\delta} - 1}{2}[1 + S(X_0) + C(X_0) - i(S(X_0) -
C(X_0))]\bigg]~,
\end{equation}
where $S(X)$ and $C(X)$ are Fresnel integral and $X_0 =
x_0/(\sqrt{\pi}R_F)$ is the reduced variable corresponding to
$x_0$. Thus, the intensity on the plane of the observer is
\begin{equation}
\label{} I_0(x_0,y_0) = A_0(x_0,y_0) A^{*}_0(x_0,y_0) = A^2
i_0(X_0)~,
\end{equation}
$i_0(X_0)$ being the relative intensity given by
\begin{equation}
\label{intrel} i_0(X_0) = 1 - (S(X_0) - C(X_0))sin(\kappa\delta) +
\bigg[S(X_0)^2 + C(X_0)^2 - \frac{1}{2}\bigg][1 -
cos(\kappa\delta)]~.
\end{equation}

In the next Section, we apply the previous formalism to the case of
monochromatic gravitational radiation (with wavelength
$\lambda_{GW}$) passing through the stellar cluster possibly
surrounding the galactic center black hole. Taking the optical path
$\delta$ to be of the same order of magnitude as the average
distance $\langle d(r)\rangle$ between the star in the cluster, we
give the condition under which Fresnel diffraction occurs. Moreover,
we shall find that a characteristic scintillation pattern appears in
the plane of the observer.

\section{Gravitational Wave Scintillation}
The Fresnel diffraction limit entails that  $D\ut>R_F$, where $D$ is
the circular hole diameter and $R_F$ is the Fresnel radius. In order
for there to be diffraction by the stellar cluster, the optical path
$\delta$ should be of the same order of $D$ that is $\simeq \langle
d \rangle$, i.e. the average distance between neighboring stars.
Therefore, we get
\begin{equation}
\label{R} \langle d \rangle \ut> \sqrt{\frac{\lambda z_0}{2\pi}}~,
\end{equation}
implying
\begin{equation}
\label{lambda} \lambda_{GW} \ut< \frac{2\pi \langle d \rangle
^2}{z_0}~.
\end{equation}
Thus, in order to have diffractive effects by the star cluster in
the Fresnel limit, we need gravitational radiation with wavelength
less than the mean spacing squared divided by the distance of the
observer from the plane of diffraction that we take to be
$z_0\simeq 8$ kpc.

Next we consider a gravitational wave source behind the stellar
cluster at the galactic center. The source might be, for example,
an extragalactic system of inspiral binary black holes or a single
spinning pulsar or a binary system of neutron stars at the outer
edge of the Galaxy, which are assumed to be detectable by a
gravitational wave telescope (like VIRGO, LIGO, LISA or ASTROD)
with an integration time $\overline{T}$ (for details see e.g.
Thorne \cite{thorne}). In the following, we shall mainly
concentrate on diffraction by the star cluster at the galactic
center and, due to the expected average distance between
neighboring stars, the most relevant gravitational radiation
sources are pulsars lying behind that cluster. Therefore, we will
consider in the following analysis the VIRGO interferometer.

\begin{figure}[htbp]
\vspace{7.cm} \includegraphics{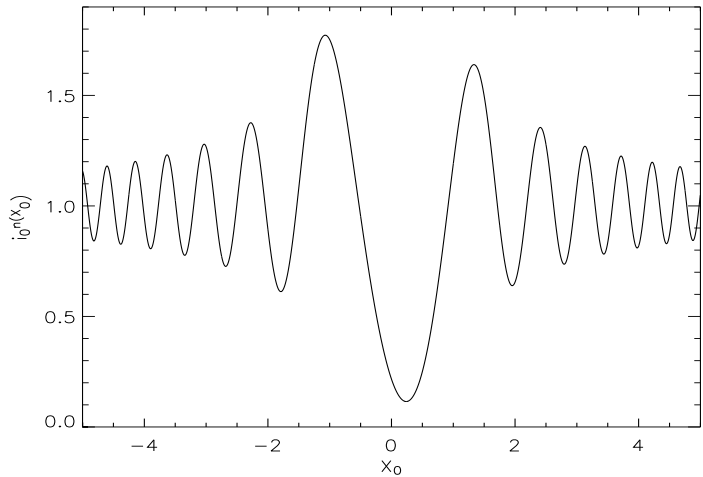} \caption{The gravitational wave pattern in
the plane of the observed is shown. Here, the diffracting stellar
cluster has parameters $\lambda_{BH}=0.99$, $\rho_0=1.15\times
10^{-11}$ g cm$^{-3}$ and $r_c=5.8$ mpc, corresponding to an
average stellar distance of $\langle d \rangle\simeq 10^{15}$ cm
evaluated at a distance of $r\simeq r_c$ from the cluster center.}
\label{f1}
\end{figure}

In Fig. \ref{f1}, we present the scintillation pattern obtained
from eq. (\ref{intrel}) in the plane of the observer for an
incoming monochromatic gravitational wave whose wavelength is
$\lambda_{GW} \simeq 1.5\times 10^7$ cm. The diffracting stellar
cluster has parameters $\lambda_{BH}=0.99$, $\rho_0=1.15\times
10^{-11}$ g cm$^{-3}$ and $r_c=5.8$ mpc, corresponding to an
average stellar distance of $\langle d \rangle\simeq 10^{15}$ cm
evaluated at a distance of $r\simeq r_c$ from the cluster center.
The position $X_0=0$ corresponds to the case of perfect alignment
among the source, the cluster center and the observer. Of course,
as one can see from Fig. \ref{f1}, as the observer moves along the
direction $X_0$, it passes through a series of maxima and minima
of the gravitational wave intensity.

\begin{figure}[htbp]
\vspace{7.cm} \includegraphics{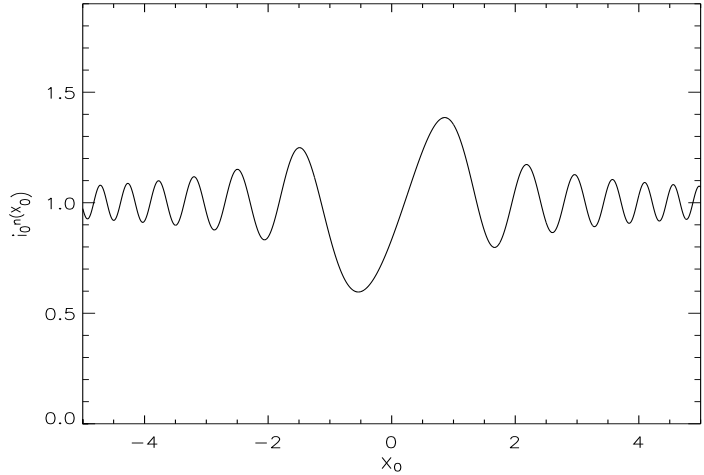} \caption{The same as in Fig. 1 but
considering, as diffracting objects, the stars with an average
distance $\langle d \rangle\simeq 10^{15}$ cm evaluated at $r\simeq
0.1 r_c$ from the cluster center.} \label{f2}
\end{figure}

Analogously to Fig. \ref{f1}, the gravitational wave pattern
obtained with $\langle d \rangle\simeq 10^{14}$ cm, evaluated at the
distance $r \simeq 0.1r_c$, is reported in Fig. \ref{f2}.

As one can see by comparing  Fig. \ref{f2} with Fig. \ref{f1}, even
though the frequency of the curves remains the same, the height of
the curves decreases and the principal maximum shifts from
$i_0\simeq 1.8$ to $i_0\simeq 1.4$. If one considers a different
gravitational radiation wavelength, which is to say a different wave
source, we obtain analogous scintillation patterns to those
presented in Fig. \ref{f1} (for $\langle d \rangle\simeq 10^{14}$
cm) or Fig. \ref{f2} (for $\langle d \rangle\simeq 10^{15}$ cm).

Obviously, the positions of the maxima do not change in the
dimensionless variable $X_0$. However, different gravitational
radiation diffraction patterns (for different incoming
wavelengths) are obtained in the variable $x_0$. As is clear, each
figure has been normalized to the incoming gravitational radiation
intensity ($I_0=1$). As such, as the observer moves, the
scintillation effect implies a change in the wave intensity that,
in principle, might be used to infer information about the source
and the cluster parameters (in particular on $\langle d \rangle$).

As an example, consider a gravitational wave (with wavelength
$\lambda_{GW} \simeq 1.5\times 10^7$ cm) interacting with the
stellar cluster at the center of the Galaxy. By inspecting Fig. 1,
it is clear that the distance between the principal maximum and the
next minimum is $\Delta X_0\simeq 1.5$. Hence, if the observer moves
around the center of the Milky Way with the typical orbital speed
$v_\odot=220 ~\rm{km~s^{-1}}$, the time-scale $\Delta T\simeq 1$
year is obtained. For such a cluster we do not consider binary
sources because of the Fresnel diffraction limits (\ref{lambda}).
Obviously, if we take into account other types of clusters (with
different physical parameters) it is also possible to get
diffractive effects with different gravitational wave sources, like
black hole binaries, etc.

In order for the diffraction effect to be measurable, it is clear
that the integration time $\overline{T}$, required by a
gravitational wave interferometer to detect a signal over the
typical instrument noise, has to be less than the traveling time,
$T_{\odot}$, for Earth to move from a maximum to the next minimum of
the scintillation pattern. Here, an optimal detection sensitivity
or, roughly speaking a signal-to-noise ratio $\simeq 1$, is assumed.
Consequently, the integration time becomes\cite{frasca}
\begin{equation}
\overline{T}=\left[\frac{h(f)}{h_0}\right]^2
\end{equation}
where $h(f)$ is the experimental sensitivity curve (in unit of
$\sqrt{Hz}$) and $h_0$ is the expected (dimensionless) gravitational
wave amplitude at the characteristic frequency $f$.

In Tab. \ref{tab1} we give the relevant parameters for three
pulsars and the resulting values for $T_{\odot}$ and
$\overline{T}$ to be evaluated by taking into account the VIRGO
sensitivity threshold \footnote{See VIR-NOT-PER-1390-51 available
online at http://www.virgo.infn.it/senscurve.}. As one can see,
for the Crab pulsar the scintillation effect is clearly
detectable, for PSR 0021-72C the effect is marginally detectable
while for PSR 1604-00 it is far from detectability, at least for
the VIRGO interferometer.

\begin{table}[htbp]
\ttbl{30pc}{Here the physical parameters for some known pulsars
together with the characteristic frequency $f$ at which the
objects mostly emit are reported. The expected gravitational wave
amplitude $h_0$ and the experimental sensitivity curve $h(f)$
(taken from VIRGO sensitivity threshold) at that frequency are
also reported. The integration time $\overline{T}$ and the
traveling time $T_\odot$ are given with the minimum average
distance $\langle d\rangle$ between two neighboring stars (see eq.
\ref{lambda}).}
{\begin{tabular}{|c|c|c|c|}\\
\multicolumn{4}{c}{} \\[6pt] \hline
Pulsar & Crab & PSR0021-72C (47TUC) & PSR1604-00  \\
\hline
$P$ (ms) & $33.2$  & $5.757$ &  $421.8$ \\
\hline
$\dot {P}$ $\times 10^{-15}$ (s/s) & $422.6$ & $0.04$ & $0.30$ \\
\hline
$f$ (Hz) & $60.423$ & $347.403$ & $4.742$ \\
\hline
$h_0$ & $3.59 \times 10^{-25}$ & $8.36 \times 10^{-27}$ & $2.68 \times 10^{-27}$ \\
\hline
$h(f)$ & $7\times10^{-23}$ & $5 \times 10^{-23}$ & $2\times10^{-20}$  \\
\hline
$T_\odot$ & $\simeq 3.75$ y &  $\simeq 1.563$ y & $\simeq 13.38$ y \\
\hline
$\overline{T}$ & $\simeq 10.6$ h & $\simeq1.1$ y & $\simeq 2\times 10^{6}$ y \\
\hline
$\langle d \rangle$ (cm) & $1.4 \times 10^{15}$ & $5.68 \times 10^{15}$ & $5.98 \times 10^{14}$\\
\hline
\end{tabular}}
\label{tab1}
\end{table}

In Tab. \ref{tab2} we give the same quantities as in Tab. \ref{tab1}
but for three different kinds of pulsar parameters: a typical radio
pulsar, a typical new-born pulsar and a millisecond pulsar. As one
can see, only for new-born pulsars, even with rather small $\dot P$
values, the scintillation effect would be detectable by the VIRGO
detector.

\begin{table}[htbp]
\ttbl{30pc}{This table is analogous to Tab. \ref{tab1} but for
three different classes of pulsar parameters.}
{\begin{tabular}{|c|c|c|c|}\\
\multicolumn{4}{c}{} \\[6pt]
\hline
$P$ (s) & $0.1$ & $0.01$ & $0.001$\\
\hline
$\dot P$ (s/s) & $10^{-15}$ & $10^{-15}$ & $10^{-19}$\\
\hline
$f$ (Hz) & $20$ & $200$ & $\simeq 2000$\\
\hline
$h_0$ & $\simeq 10^{-26}$ & $\simeq 3\times 10^{-26}$ & $\simeq 10^{-27}$\\
\hline
$h(f)$ & $5\times10^{-22}$ & $4\times 10^{-23}$ & $10^{-22}$\\
\hline
$T_\odot$ & $\simeq 6.5$ y & $\simeq 2$ y & $\simeq 0.92$ y\\
\hline
$\overline{T}$ & $\simeq79$ y & $\simeq 18.5$ d & $\simeq 300$ y\\
\hline
$\langle d\rangle$ & $2.42 \times 10^{15}$ & $7.66 \times 10^{14}$ & $2.42 \times 10^{14}$\\
\hline
\end{tabular}}
\label{tab2}
\end{table}

\section{Concluding Remarks}

The General Theory of Relativity predicts the existence of
gravitational waves. In accordance with what is currently observed
for electromagnetic waves interacting with obstacles (or slits), we
expect that also gravitational radiation may undergo diffraction. As
usual, it is possible to have diffraction effects only if the
incident radiation wavelength and the intervening obstacle size are
of the same order of magnitude. Hence, due to the typical large
wavelength of the considered radiation (i.e. $10^{7}-10^{15}$ cm for
spinning neutron stars), only systems of objects on galactic scale
may act as diffractive objects. We have considered the diffraction
of gravitational waves by a stellar cluster possibly hosted at the
center of our Galaxy and found, under the conditions for which the
Fresnel diffraction occurs, that a {\it scintillation} of the
gravitational wave signal comes out naturally.

As we have shown, this effect, analogous to the well known
scintillation of radio waves due to a medium along the line of
sight of a distant source, produces a characteristic pattern on
the plane of the observer. We have also shown that the typical
crossing time $T_{\odot}$ (i.e. the time required by the observer
to move across a minimum to the next maximum of the gravitational
wave scintillation pattern) is of the order of a few years. For a
particularly strong gravitational wave signal and in particular
regions of the frequency band, the typical integration time of
instruments like VIRGO and/or LIGO (which have a rather similar
sensitivity threshold curve) may be as short as few hours thus
allowing a good sampling of the scintillation pattern.

We emphasize that, as observed in Section 4, new-born pulsars,
i.e. pulsars that were born within the last $\sim 10^4$ years are
the best candidates to look for the gravitational wave
scintillation effect by the star cluster lying at the galactic
center with the VIRGO instrument.

It is worth noting that the next generation of gravitational wave
interferometers (such as Advanced VIRGO and Advanced LIGO) will have
a sensitivity greater than that allowed by the present instrument
configurations. For example, Advanced LIGO \footnote{See the web
page at http://www.ligo.caltech.edu/advLIGO/.} will have more than
an order of magnitude greater sensitivity than initial LIGO implying
a reduction by at least two orders of magnitude of the integration
time. If this is the case, there will be a real possibility of using
the gravitational radiation signals to probe deep into our Universe
and search for effects like the scintillation of gravitational
waves. Detecting this effect will give an independent way of
estimating the main parameters of the intervening stellar cluster
like the average stellar distance $\langle d\rangle$ and the cluster
central mass density. The microlensing effect on the gravitational
waves due to the black hole at the galactic center has been
investigated elsewhere \cite{dinq}.



\begin{thebibliography}{0}

\bibitem{stm} T. Suyama, R. Takahashi and S. Michikoshi, {\it Phys. Rev.} D, {\bf
72}, 043001 (2005).

\bibitem{tsm} R. Takahashi, T. Suyama and S. Michikoshi, {\it Astron. and
Astrophys.}, {\bf 438}, L5 (2005).

\bibitem{macquart} J. P. Macquart, {\it Astron. and Astrophys.}, {\bf 422}, 761 (2004).

\bibitem{ruffa} A. A. Ruffa, {\it Astrophys. J.} {\bf 517}, L31 (1999).

\bibitem{dinq} F. De Paolis et al., {\it Astron. and Astrophys.}, {\bf 394}, 1065
(2001).

\bibitem{din} F. De Paolis et al., {\it Astron. and Astrophys.}, {\bf 366}, 749
(2002).

\bibitem{bre} G. Congedo et al., {\it Proceedings of the Second International
ASTROD Symp., June 2-3, 2005, Bremen},
in press on Gen. Rel. Grav., Springer Verlag (2006).

\bibitem{Ghez04} A.M. Ghez et al., ApJL, {\bf 601}, L159, (2004).

\bibitem{Ghez05} A.M. Ghez et al., ApJ, {\bf 620}, 744, (2005).

\bibitem{Ghez03} A.M. Ghez et al., ApJL, {\bf 586}, L127, (2003).

\bibitem{born} J. Born and S. Wolf, {\it Principles of Optics}, Cambridge
University Press, (1999).

\bibitem{moniez} M. Moniez,  {\it Astron. and Astrophys.}, {\bf 366}, 1065,
(2001).

\bibitem{frasca} S. Frasca et al.,  {\it Class. Quantum Grav.}, {\bf 22}, S1013,
(2005).

\bibitem{s2} F. De Paolis et al., {\it New Astronomy}, in press
(2006).

\bibitem{vetrugno} D. Vetrugno, {\it Thesis} at the University of Lecce, December
2005, available online at
http://www.fisica.unile.it/$\sim$nucita/.

\bibitem{thorne} K. S. Thorne, {\it Reviews in Modern Astronomy}, {\bf 10}, 1 (1997).

\end{thebibliography}
\end{document}